\documentclass[useAMS,usenatbib]{mn2e}
\usepackage{graphicx}

\def\hmpc{h^{-1}{\rm Mpc}}

\def\hmsun{h^{-1}M_\odot}
\def\lcdm{$\Lambda$CDM}

\newcommand{\beq}{\begin{equation}}
\newcommand{\eeq}{\end{equation}}
\newcommand{\beqa}{\begin{eqnarray}}
\newcommand{\eeqa}{\end{eqnarray}}
\newcommand{\om}{\Omega_m}

\newcommand{\dl}{\delta}
\newcommand{\dhh}{\delta H^2}

\title[Cosmic Structure Growth and Dark Energy]{Cosmic Structure Growth 
and Dark Energy}
\author[Linder \& Jenkins]{E.V. Linder$^{1}$\thanks{E-mail:
EVLinder@lbl.gov; A.R.Jenkins@durham.ac.uk},
A. Jenkins$^{2}$ \\
$^{1}$ Physics Division, Lawrence Berkeley National Laboratory,  Berkeley, California 94720, USA\\
$^{2}$Institute for Computational Cosmology, Physics Department, Durham University, South Road, Durham DH1 3LE,
UK\\
}
\begin{document}

\date{}

\pagerange{\pageref{firstpage}--\pageref{lastpage}} \pubyear{2002}

\maketitle

\label{firstpage}

\begin{abstract}
Dark energy has a dramatic effect on the dynamics of the universe,
causing the recently discovered acceleration of the expansion.  The
dynamics are also central to the behavior of the growth of large scale
structure, offering the possibility that observations of structure
formation provide a sensitive probe of the cosmology and dark energy
characteristics.  In particular, dark energy with a time varying
equation of state can have an influence on structure formation
stretching back well into the matter dominated epoch.  We analyze this
impact, first calculating the linear perturbation results, including
those for weak gravitational lensing. These dynamical models possess
definite observable differences from constant equation of state
models.  Then we present a large scale numerical simulation of
structure formation, including the largest volume to date involving a
time varying equation of state. We find the halo mass function is well
described by the Jenkins et al mass function formula.  We also show how to
interpret modifications of the Friedmann equation in terms of a time
variable equation of state.  The results presented here provide 
steps toward realistic computation of the effect of dark energy
in cosmological probes involving large scale structure, such as
cluster counts, Sunyaev-Zel'dovich effect, or weak gravitational
lensing.
\end{abstract}


\section{Introduction }\label{sec.intro} 

Direct dynamical measurement of the expansion of the universe through the 
Type Ia supernova distance-redshift method discovered that the expansion 
is accelerating \citep{perl99,riess98}.  This has wide reaching implications 
with respect to the fate of the universe, its dominant constituent, 
and the nature of fundamental physics.  Some 70\% of 
the total energy density acts like a dark energy with strongly 
negative pressure.  Subsequent observations of the cosmic 
microwave background (CMB) power spectrum and of large scale structure 
give a concordant picture \citep{bond,perc,sper}.  

Mapping the expansion history of the universe offers a way to gain insights 
into the mysterious dark energy.  For example, characterizing its  
equation of state (pressure to energy density ratio) behavior gives 
insight on the properties of the high energy physics scalar field 
potential.  Distance measures, notably the supernova method, have proved 
adept at beginning to constrain the energy density and equation of state 
of the dark energy.  Recent limits, combined with CMB or large scale 
structure information, within the low redshift approximation 
of a constant equation of state (EOS) ratio $w$, give $-1.61<w<-0.78$ at 
95\% confidence 
\citep{knop}.  Great improvements should occur in the next decade 
with, e.g., the Supernova/Acceleration Probe \citep[SNAP; ][]{snap}
dedicated dark energy program, in particular extending to 
constraints on the generically expected time varying function $w(z)$.  

Several other cosmological probes look promising, though their
systematic uncertainties are less well defined.  But the eventual
synergy of independent and complementary methods should prove powerful
in revealing the nature of dark energy.  Some of these probes offer
the opportunity to measure fairly directly the expansion rate behavior
$H(z)$, rather than just the distance which involves this quantity
through a redshift integral.  In turn, $H(z)$ involves an integral
over the equation of state $w(z)$.  One cannot however na{\"\i}vely
assume that a limited measure of $H(z)$ is better than knowing the
distance over a wide redshift range.  This was demonstrated for the
cosmic shear, or Alcock-Paczy{\'n}ski, effect in \cite{alpac} and the
baryon oscillation probe in \cite{bosc} \citep[see][for a variety of
other methods]{huttur}.  But such information can prove valuable in
complement with precision distance data.

Furthermore, methods involving the growth of large scale structure appear, 
initially at least, to possess sensitivity to the cosmic equation of state. 
This enters through the actual growth of density fluctuations, 
i.e.~the balance of attractive gravitational stability with the dynamic 
friction of the expansion, through its evolution with redshift or time as 
the dynamics changes under the influence of dark energy, and through the 
cosmic volume available in which structures form.  However little rigorous 
work has been done on the full impact of dark energy other than for the time 
independent cosmological constant model.  This especially 
applies to the virtually universal time varying EOS models, where not just 
the energy density but the function $w(z)$ is time dependent.  Indeed a 
constant $w$ model (other than the cosmological constant $w=-1$) functions 
only as a crude approximation, unsuitable for next generation data that 
extends beyond $z\approx0.5$ or that seeks to combine complementary probes.  
Moreover, the important and revelatory physics responsible for 
the accelerating 
universe appears in the field dynamics -- the time variation $w'\sim dw/dz$. 

If we desire to take advantage of the power of large scale structure formation 
as a probe of dark energy, we must include sufficient realism in the model 
that we form a consistent picture of the underlying physics.  This includes 
both systematic uncertainties in the astrophysics and observations, and 
a treatment of time variation in the dark energy EOS.  Structure based 
methods such as weak lensing, Sunyaev-Zel'dovich distance measures, and 
cluster counts all require knowledge of how structure formation behaves in 
the presence of realistic dark energy.  In Section \ref{sec.wp} we present the 
key role of $w'$ in describing the dynamics of the expansion and in 
Section \ref{sec.linpert} solve the 
growth equation in linear perturbation theory.  Section \ref{sec.sim} 
describes 
the numerical simulation of large scale structure in the presence of dark 
energy and Section \ref{sec.res} discusses the results.  Implications and an 
outline of future research are presented in Section \ref{sec.concl}.

\section{{\lowercase{$w'$}} Is Everywhere }\label{sec.wp}

The expansion rate of the universe, $H(z)=\dot a/a$, where $a(t)$ is the 
scale factor, enters into both the kinematics and dynamics of the 
cosmological model.  Distances are integrals of the proper or conformal  
time; for example in a flat universe (assumed throughout) the comoving 
distance is 
\beq
r(z)=\int_a^1 dt/a(t)=\int_0^z dz/H(z), 
\eeq
where the redshift $z=a^{-1}-1$.  The angular diameter distance is just 
$r_a=r/(1+z)$ and the luminosity distance is $r_l=(1+z)\,r$.  The expansion 
rate, or Hubble parameter, $H(z)$ also enters into dynamical quantities 
such as the growth of structure through a ``Hubble drag'' term. 

Within the flat universe, dark energy picture, the Friedmann equations 
give the expansion rate as 
\beq
H^2(z)/H_0^2 = \Omega_m(1+z)^3 + (1-\Omega_m)\,
e^{3\int_0^z d\ln(1+z')[1+w(z')]},
\label{fried}\eeq
where $H_0$ is the Hubble constant, the present value of the Hubble 
parameter, $\om$ is the dimensionless matter density today (so the dark 
energy density is $1-\om$), and $w(z)$ is the generically time dependent dark 
energy equation of state.  While each model of dark energy has a particular 
form for $w(z)$, in order to compare models one usually adopts a 
parametrisation.  The one proposed by \cite{lin0208}: $w(z)=w_0+w_az/(1+z)$,  
with $w_0$ the present value of the EOS and the time variation 
$w'\equiv dw/d\ln (1+z)|_{z=1}=w_a/2$, allows consideration of data extending 
to $z>1$ and presents an excellent approximation to slow roll scalar field 
dark energy models \citep{lin0210}.  Thus we can use the observations 
in a well defined manner to investigate the fundamental physics manifesting 
in the EOS time variation. 

However, a time varying equation of state is more general than this.  Some 
theories have been proposed that explain the acceleration not through a 
scalar field dark energy but through modifications of the Friedmann 
equations themselves by alternative theories of gravitation, e.g.\ arising 
from extra dimensions, or by highly speculative components such as 
the Chaplygin gas or quantum or higher dimensional corrections.  In 
\cite{lin0208}, a formalism was presented to use supernova distance data to 
constrain these scenarios by means of mapping the expansion history $a(t)$ 
directly, rather than using an intermediate parametrisation $w(z)$.  While 
this is valid, and probably physically preferred, we note here an alternate 
interpretation. 

Consider Eq.\ (\ref{fried}).  The dark energy term really just describes our 
ignorance concerning the physical mechanism leading to the observed effects 
of acceleration in the expansion, i.e.\ an increase in the expansion rate. 
Let us instead write this as 
\beq 
H^2(z)/H_0^2 = \Omega_m(1+z)^3 + \dhh/H_0^2, \label{dh} 
\eeq 
where now we encapsulate {\it any} modification to the Friedmann equation 
of general relativity in the last term.  That is, we take a very empirical 
approach: all we have observed 
for sure is a certain energy density due to matter, $\om$, and consequences 
of the expansion rate $H(z)$. 

We can now write the deceleration parameter generally as 
\beq 
q\equiv -{a\ddot a\over \dot a^2} = {1\over2}-{3\over2}{\dhh\over H^2}- 
{1\over2}{\dot{(\dhh)\over H^3}}. \label{qdh} 
\eeq 
If we interpret the modified expansion rate as being due to a $w(z)$ 
as appearing in Eq.\ (\ref{fried}) -- whether or not it has anything to 
do with a scalar field -- then we find 
\beq 
w(z)\equiv -1+{1\over3}{d\ln\dhh\over d\ln(1+z)}\,. \label{wdh} 
\eeq 
This now defines an effective, time varying equation of state 
\citep[something similar was noticed by][]{staro}.  
Of course it reduces to the usual result in the scalar field dark 
energy case.  But this goes to illustrate the centrality of a 
time varying $w(z)$ -- or something that looks just like it -- 
for probing cosmological models.  To fix the EOS to be a constant 
$w$ rather than a varying function $w(z)$ is highly non-generic, 
an unjustified assumption and a frequently poor approximation, and 
can blind us to important physics. 

\section{Linear Perturbation Theory} \label{sec.linpert}

The growth of structure depends sensitively on the expansion rate 
of the universe.  For example, the solution to the classic Jeans 
instability in a static space shows exponential growth under 
gravity, while this gets reduced to only a power law behavior in 
time in an expanding space-time.  The perturbations are sourced 
in the gravitational instability of slightly denser regions having 
correspondingly greater gravitational attractions and thus 
further increasing in density; this is opposed by an effective 
friction, or ``Hubble drag'' term, due to the expansion. 

Since dark energy affects the expansion rate, one expects to see 
an influence due to the friction.  That this can be substantial 
comes from our experience that open universes, which effectively 
have a component with EOS $w=-1/3$, can shut off the growth of 
structure when the curvature energy density dominates over the 
matter density.  Similar results hold for a cosmological constant 
dominated universe.  Thus for 
general dark energy models the state of structure formation at 
various redshifts could probe the equation of state, and even its 
time variation (we fix the sound speed to its canonical value, $c_s=1$).  
We first present a general, pedagogical analysis 
of the influence of EOS on growth in the linear perturbation regime, 
then specifically calculate the results for various dark 
energy models. 

\subsection{Growth of Linear Perturbations}\label{lingrow}

On scales smaller than the horizon, the dark energy component is 
expected to be smooth \citep{ma,dave} so we only consider 
perturbations to the matter.  Then the growth equation becomes 
\beqa  
\ddot\dl +2H\dot\dl -(3/2)H^2\Omega_m\dl &=& 0 \\ 
\dl''+(2-q)a^{-1}\dl'-(3/2)\Omega_ma^{-2}\dl &=& 0, 
\eeqa 
where $\dl$ is the fractional matter density perturbation, $q$ 
is the deceleration parameter, dot denotes a time 
derivative and prime a derivative with respect to scale factor $a$.  
One can readily see that growth in a universe with $\Omega_{tot}$ 
and $\Omega_{-1/3}$ in a component characterized by $w=-1/3$ behaves 
like growth in a universe with no such component but 
$\Omega_{tot}'=\Omega_{tot} -\Omega_{-1/3}$.  So a flat universe 
with a $w=-1/3$ component acts like an open universe. 

We can write the growth equation in terms of the general EOS $w(z)$, 
which as we saw in \S\ref{sec.wp} can also represent modifications 
to the framework of the theory. Because the interpretation of the 
source and drag terms is so straightforward, this generalization 
broadly (but not always: see the object lesson within standard 
gravitation in the Appendix) carries over to the growth dynamics.  
Defining the growth as the ratio of the perturbation 
amplitude at some scale factor relative to some initial scale 
factor, $D=\dl(a)/\dl(a_i)$, the equation becomes \citep{mpa} 
\beqa 
D'' + {3\over2}\left[1-{w(a)\over 1+X(a)}\right]\,{D'\over a}-{3\over2} 
\,{X(a)\over 1+X(a)}\,{D\over a^2}=0\,. \label{da} 
\eeqa
\beqa 
X(a) &=& {\om\over 1-\om} e^{-3\int_a^1 d\ln a'\,w(a')} \label{eq.xw}\\ 
&=& \om a^{-3}\Big/(\dhh/H_0^2) \label{eq.xh}
\eeqa 
where Eq.\ \ref{eq.xw} gives the general case of $X$ and Eq.\ \ref{eq.xh} 
puts it in terms of the time dependent scalar field 
equation of state. 

The variable $X$ is the ratio of the matter density to the dark 
energy density, and the growth equation holds even if flatness does not, 
so we do not set $X/(1+X)=\om(a)$.  As $X$ gets small the source term 
vanishes and growth cannot generally proceed.  Note that this is the 
reason why (subhorizon dark matter) perturbations cannot effectively 
grow in the radiation dominated epoch, not from any overriding 
influence of the Hubble drag term.  (Of course we have not included 
coupling between components; radiation-baryon coupling prevents 
gravitational instability in the baryon 
perturbations). From Eq.\ (\ref{da}), 
one sees that the friction term opposing the growth is 
proportional to $1-w$ when matter is not dominant.  So in the 
radiation epoch the drag is less than in an open or accelerating 
epoch.  It is only for those cases when $w<1/3$, as comes from 
solving the characteristic equation, that the 
growth is shut off by the friction term.  (Indeed growth can 
occur for $w>1/3$.) 

One can readily verify that for large $X$ one 
recovers the matter dominated behavior $D\sim a$.  It is convenient 
to remove this trend, common to all the considered cosmological 
models at high redshift, and define a variable $G=D/a$.  The 
evolution equation for this ``normalized'' growth is 
\beq 
G''+\left[{7\over2}-{3\over2}{w(a)\over 1+X(a)}\right]\,{G'\over a} 
+{3\over2}\,{1-w(a)\over 1+X(a)}\,{G\over a^2}=0\,. 
\label{eq.ga}\eeq 
Through the cosmological Poisson equation $G$ is related to 
both the gravitational potential and peculiar velocity fields. 

\begin{figure}
\includegraphics[width=8cm]{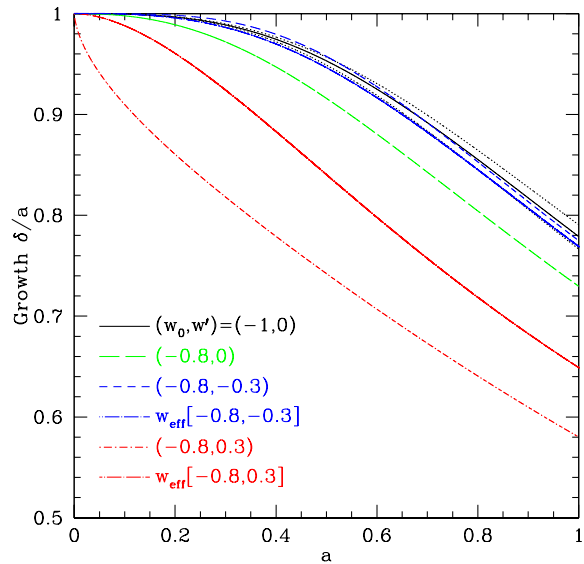}
\caption{Growth factor of linear density perturbations plotted 
vs.\ scale factor $a=(1+z)^{-1}$ for six dark energy models with 
$\om=0.3$.  Dotted curves show the effect of changing $\om$ by 
$\pm0.02$ for the cosmological constant case.  Curves labeled 
$w_{eff}$ use the constant equation of state that gives the same 
distance to the last scattering surface as the bracketed models, 
and hence are degenerate with respect to the CMB.}
\label{fig.lingrow}
\end{figure}

This formalism readily allows incorporation of dark energy with 
a time varying EOS, or the alternate equation of motion 
Eq.\ (\ref{dh}).  Figure \ref{fig.lingrow} illustrates the 
solutions 
for a variety of dark energy models.  Note that models with 
constant equation of state have modest differences from the 
cosmological constant case, $\sim 7\%$ for $w=-0.8$.  This is 
still larger than the differences caused by changing the matter 
density by $\pm0.02$, shown by the dotted curves flanking the 
cosmological constant model. 

Dark energy whose EOS is negatively evolving, $w'<0$, e.g.\ 
acts more like a cosmological constant in the past but with 
a less negative EOS today, is almost indistinguishable in its 
linear growth predictions from the cosmological constant.  This 
is because the dark energy contribution to the total energy 
density only becomes significant at late times in these models. 
This property holds as well for dark energy models with 
EOS $w<-1$. 

However, the class of models with $w'>0$ exhibits dramatically 
different behavior.  This includes what we chose as our fiducial 
model to test time varying EOS, the supergravity inspired model 
SUGRA of \cite{braxm} that is well fit by $w_0=-0.82$, $w_a=0.58$ 
(i.e.\ $w'=+0.29$).  Note that \cite{lin0210} has shown that the 
fit is good to 3\% in $w$ at $z=1.7$ and to 0.2\% in distance 
to the last scattering surface at $z=1089$. 

Models with $w'>0$ show definite differences in the growth of 
structure from a cosmological constant universe.  For a model 
like SUGRA (here actually $w_0=-0.8$, $w'=0.3$) the growth disparity reaches 
26\% at the present day.  The figure shows two other points of 
interest.  Comparing SUGRA to the constant EOS model that 
matches its distance to the last scattering surface, 
and so effectively mimics the time varying model as far as CMB 
data is concerned (except at low multipoles), the difference 
in total growth is 12\%.  That is, 
the time varying model and the constant, effective model can 
be distinguished through large scale structure information, 
despite being largely degenerate in their CMB power spectra.  
This is a promising sign. 

The other interesting element is that even at high redshift the 
time varying model with $w'>0$ possesses different behavior from 
the cosmological constant and other constant $w$ models.  This shows 
the influence of dark energy at early times is very different from 
the cosmological constant, which quickly becomes dynamically 
negligible for $z>1$.  Such a characteristic offers the 
possibility that evidence for time varying EOS dark energy might 
be found as well in the low multipole region of the CMB power 
spectrum \citep[see][for a discussion of the influence of 
``early quintessence'' on the CMB through the integrated Sachs-Wolfe 
effect]{caldwell}. 

So three avenues appear to be open for the detection of physically 
important properties of dark energy from its influence on large 
scale structure: 1) linear growth rate, 2) nonlinear structure 
formation and evolution, and 3) large angle power in the CMB power 
spectrum.  In the remainder of this section we investigate further 
the first avenue and then proceed in \S\ref{sec.sim} to discuss 
numerical simulations of 
the second possibility \citep[see][for an analytic attempt]{benabed}.  
The third approach has been addressed by 
\cite{isw} with 
not very optimistic conclusions \citep[but see][]{coodrag}.  

\begin{figure}
\includegraphics[width=8cm]{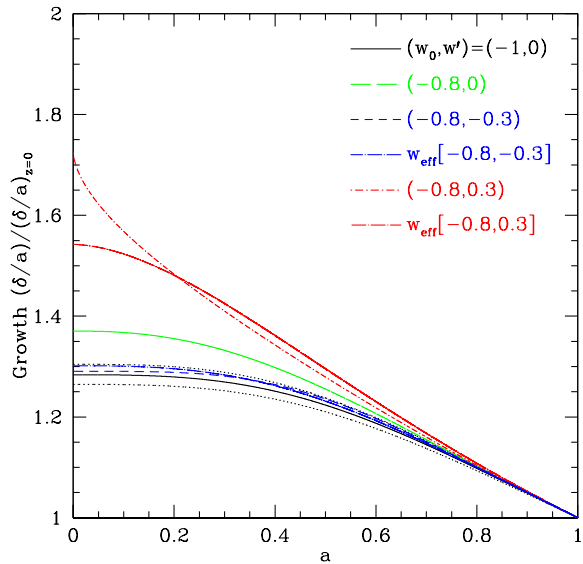}
\caption{Same as Fig.\ \ref{fig.lingrow} but with the growth 
factors normalized by their values today.  The present growth, 
related to the mass variance $\sigma_8$, is not yet precisely 
known, however. 
} 
\label{fig.lingrow.z0}
\end{figure}

Somewhat different results for the linear growth rate appear if 
we change the measurement technique.  Suppose that rather than 
normalizing the density perturbations by their high redshift behavior 
(currently corresponding to COBE/WMAP normalisation of the matter power 
spectrum, though Planck will provide great improvement), we calibrate 
them by their present amplitude.  This is like 
fixing $\sigma_8$, the power on the interface of linear/nonlinear 
scales.  While observations have not yet determined this precisely, 
we can explore the consequences.  As seen in Fig.\ 
\ref{fig.lingrow.z0}, now the models are difficult to distinguish 
at low redshifts and there is near degeneracy in the growth 
factor between the time varying model and the effective constant 
$w$ model that matches it with respect to the CMB.

The issue of the optimal way to measure the growth factor through 
the matter power spectrum is an area of ongoing research, increasingly 
important with the future advanced large scale structure surveys. 
Perhaps the most realistic observable for now is the ratio of growth factors 
at different redshifts, a measure of the evolution of structure. 
This will carry correspondingly less content since it is only 
a relative measure, without information from the absolute level. 
Such a ratio could be read off from either Fig.\ \ref{fig.lingrow} 
or \ref{fig.lingrow.z0} and indeed can be seen to vary little with 
cosmological model.  For example, the evolution between $z=2$ and 
$z=1$ in the models with $(w_0,w_a)=(-0.8,0)$ or $(-0.8,0.3)$ agrees 
with the evolution in the cosmological constant model to 2\% or 5\%. 
This casts strong doubt on the idea that the growth evolution of 
large scale structure by itself (without a precise and robust high 
redshift, CMB normalisation) is useful as a probe of dark energy. 

\subsection{Sensitivity of Linear Growth Rate to Dark Energy 
\label{growsens}} 

\begin{figure}
\includegraphics[width=8cm]{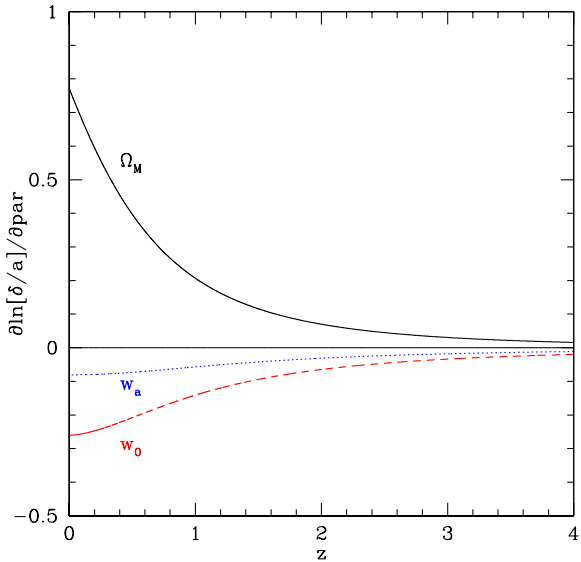}
\caption{Logarithmic sensitivity of the growth factor to the 
cosmological parameters, perturbed around the cosmological 
constant model, as a function of redshift.  Such a plot is 
useful in finding lower bounds to parameter estimation errors 
and in indicating degeneracies between parameters. 
} 
\label{fig.ddl}
\end{figure}

To understand the use of the evolution of linear matter density 
perturbations for probing the dark energy, we need to study not 
just the gross differences in the curves in Fig.\ \ref{fig.lingrow} 
but the details of how they depend on dark energy properties 
and other cosmological parameters.  We use the Fisher matrix 
formalism \citep[see, e.g.,][]{tegfis} to plot the sensitivity 
of the growth to the parameters $\om$, $w_0$, and $w_a$ in 
Fig.\ \ref{fig.ddl}.  The sensitivity increases toward low 
redshift since this corresponds to more time for the differing 
growth dynamics to take effect. 

Besides the sensitivity, the degeneracy between parameters is 
a crucial aspect to the usefulness of the probe.  The similarity 
of the shapes of the $w_0$ and $w_a$ curves guarantees a strong 
degeneracy between them, apart from further interaction with the 
value of $\om$.  Combined with the relatively low accuracy 
achievable on the growth factor from observations, this leads 
to the growth factor by itself -- even with the initial amplitude known -- 
not serving as a precision 
cosmological probe.  For example, even if all other parameters were 
fixed, a 5\% determination of $\delta/a$ at $z=1$, say, would only 
constrain $w_0$ to $\pm0.35$.  And degeneracies strongly amplify the 
uncertainty.  Additionally, the degeneracy direction in the $w_0-w_a$ 
plane is roughly aligned with the contours from the CMB, so the growth 
factor possesses little complementarity with the CMB and only a 
modest amount with the supernova distance measure. 

\begin{figure}
\includegraphics[width=8cm]{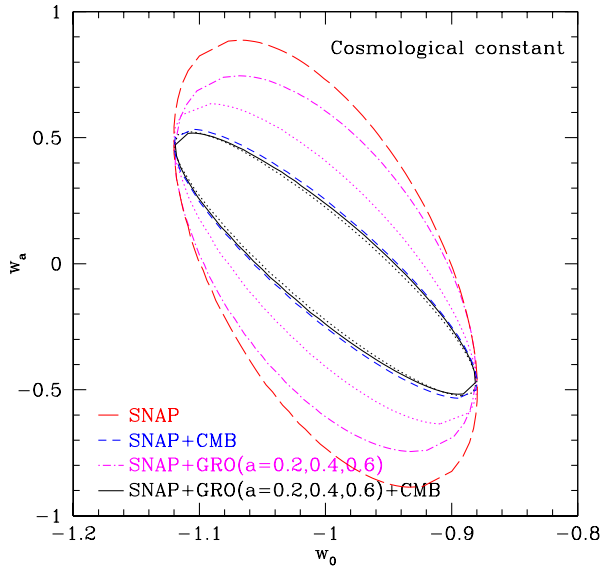}
\caption{Parameter estimations (68\% confidence level) of the present 
equation of state and its time variation, marginalizing over the 
matter density $\om$ with a prior of $\sigma(\om)=0.03$.  Dotted 
curves use the growth factor normalized to its value today.  The 
growth factor alone (not shown) is poor in constraining the 
cosmological model, and offers little complementary leverage for the 
proposed SNAP supernova 
survey, and none in addition to SNAP plus the Planck CMB survey. 
} 
\label{fig.sngr}
\end{figure}

For the simulated data sets we consider a measurement of the 
growth factor to 5\% (equivalent to 10\% 
determination of the linear power spectrum) at three values of the 
scale factor (equivalent to $z=0.67$, 1.5, 4), SNAP supernova 
measurements ({\it including} systematic uncertainties), and Planck 
CMB determination of the angular distance to the last scattering 
surface (normalisation of the primordial power is implicit in 
the growth factor measurement).  The results, shown in 
Fig.\ \ref{fig.sngr}, are not sensitive to the exact redshifts 
chosen for the growth factor estimates.  Addition of growth information 
improves constraint of $w_0$, $w_a$ by less than 2\%.  This holds 
as well if the growth factor is normalized to $z=0$.  

\begin{figure}
\includegraphics[width=8cm]{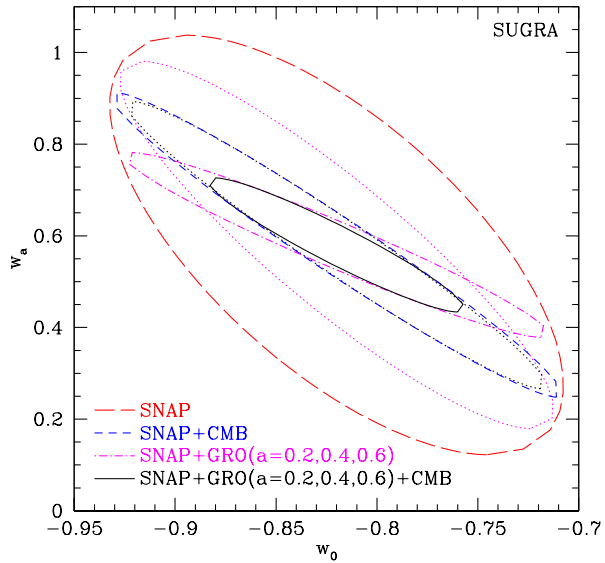}
\caption{Same as Fig.\ \ref{fig.sngr} but for dark energy 
following the SUGRA model, with present equation of state 
$w_0=-0.82$ and time variation $w_a=0.58$.  In this case, 
information on the linear growth factor adds valuable constraints. 
} 
\label{fig.sngr.sug}
\end{figure}

However, 
as was found for the baryon oscillation method \citep{bosc}, the 
situation changes dramatically when the fiducial dark energy model 
is not taken to be the cosmological constant but the SUGRA model 
with time varying EOS.  Now the growth factor under the influence 
of the dynamic field has a different cosmological dependence and 
possesses substantial complementarity with both the supernova and 
CMB data, as shown in Fig.\ \ref{fig.sngr.sug}.  The parameter 
uncertainties using all three methods in synergy become 
$\sigma(w_0)=0.04$, $\sigma(w')=0.05$ (recall that $w'=w_a/2$). 
That gives a 6$\sigma$ detection of time variation in the EOS! 

A further interesting characteristic of the growth factor is a 
sufficiently distinct cosmological dynamics dependence to break 
some degeneracies outside 
the dark energy framework.  For example, the braneworld model 
with crossover scale $H_0r_c=1.43$ discussed in \cite{lin0208}, 
difficult to distinguish from a $w=-0.7$ dark energy model through 
the distance relation ($<0.5\%$ difference), does differ by 
$\sim4\%$ in the 
growth factor (though not when normalized at low redshift).  This 
illustrates the utility of the $\dhh$ formalism of \S\ref{sec.wp} for 
testing the cosmological framework. 

\subsection{Growth Rate and Weak Gravitational Lensing}\label{sec.wl}

One of the main applications of the growth rate is to cosmological 
observables that statistically characterize large scale structure.  In 
\S\ref{sec.sim} we address how the ingredient of the linear growth 
factor enters description of nonlinear structure.  But here we make 
a brief, illustrative foray into the linear regime of the weak 
lensing shear used to map out the matter distribution (including 
dark matter) through its gravitational 
deflection of light from distant sources.  

Observations of such 
shear are becoming increasingly useful cosmological tools 
and depend on the primordial matter density 
power spectrum, the growth rate, and geometric distance factors. 
We examine the simplified, though central, quantity of the shear 
linear lensing power spectrum \citep[cf.][]{jainsel,huterer}: 
\beq 
C_\ell\sim \int_0^{z_s} dz_l\,\left[\om(1+z_l)\left(r_{ls}\over 
r_s\right) \,G(z_l)\right]^2{1\over H(z_l)\,r_l}\,.\label{wl}
\eeq 
Here $z_l$ is the 
redshift of the lensing mass, $z_s$ of the source galaxy, $r_s$ the 
comoving distance to the source, and $r_{ls}$ the comoving distance 
between the source and the lens.  For simplicity we used the linear 
matter power spectrum $P_k\sim k\,G^2$ and fix $z_s$.  The latter 
corresponds to having tomographic information. 

\begin{figure}
\includegraphics[width=8cm]{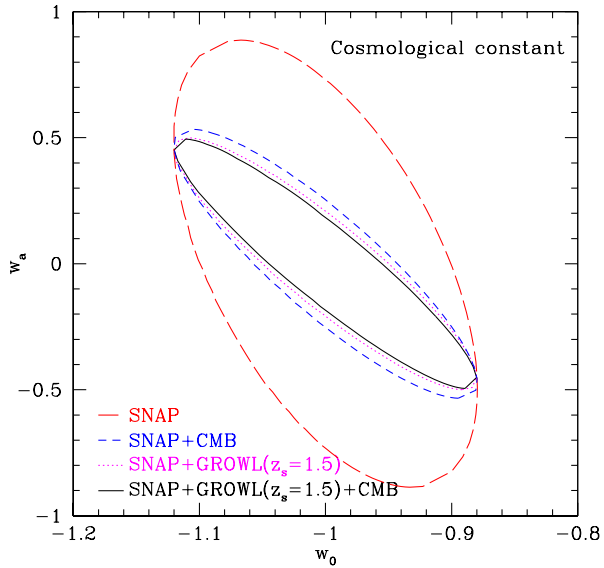}
\caption{Parameter estimations (68\% confidence level) of dark energy 
properties using future 5\% measurement of the weak gravitational 
lensing linear power spectrum in addition to SNAP supernova distances and 
Planck CMB data. The lensing power has good complementarity with 
supernovae, though little with the CMB.  Thus supernovae plus lensing 
should provide a valuable crosscheck on supernovae plus CMB results. 
} 
\label{fig.sngrowl}
\end{figure}

We find that this lensing combination alone cannot place useful 
constraints on the dark energy but it does have good complementarity 
with the supernova distance measurements.  Such lensing data strongly 
improves the determination of $\om$, as traditionally expected, and reduces 
uncertainty in the time variation $w_a$ slightly more dramatically than 
CMB information.  However it has little complementarity with CMB data, 
and so the limits are not much further improved (1\%) on inclusion 
of both lensing and CMB data.  Fig.\ \ref{fig.sngrowl} shows the case 
of 5\% determination of the weak lensing shear power spectrum at $z_s=1.5$.  
Results are not very sensitive to the exact 
redshift, or combination, chosen.  The SUGRA 
case leads to similar conclusions, though there the weak lensing has 
more complementarity with the CMB and adding the lensing information 
tightens the constraints on $w_0$ and $w_a$ from supernovae and CMB by 
34\%.  Thus the SNAP weak lensing program 
in combination with the SNAP supernova survey provides an important 
crosscheck on the results from SNAP supernovae plus CMB, plus the 
possibility of additional improvements. 

Of course the probe may be further strengthened by using information 
from the nonlinear part of the power spectrum.  To employ the 
nonlinear regime, one needs 
to carry out numerical simulations including a time varying EOS or use 
fitting formulas valid for such cases, neither of which previously 
existed.  In the remainder of the paper we present and discuss such 
a simulation.

\section{A Simulation of Large Scale Structure Formation} \label{sec.sim}

  So far in this paper we have concentrated on the dependence of the
linear growth factor on the nature of the dark energy.  In this
section we turn our attention to the influence of the dark energy on
the growth of non-linear large-scale structure -- specifically the 
dark matter halo mass function.  This measure of the number density 
of objects as a function of mass is important in a wide range 
of large-scale structure areas, including the use as a probe of the 
cosmological model. 

  Theoretically the dark matter halo mass function has been explored
in CDM models mainly through n-body simulations
e.g. \cite{EFWD88,LC94, GROSS98,Gov99,J01,White01,White02}. Nearly all
simulations to date focussing on this issue have modelled a universe
in which the dark energy is static -- a cosmolological constant
$\Lambda$ -- and the matter component is exclusively dark -- and have
ignored the fluid and dissipative nature of the baryonic component.
The baryons are taken into account when computing the input matter
transfer function used in making the initial conditions for the
simulation.  In the simulation itself it is assumed that the baryons
behave as a pressureless component which mirrors the dark matter
precisely.  Modelling of the formation of individual haloes where a
baryonic component is included, for example the Santa Barbara cluster
comparison project \citep{SB99}, lends support to the idea that
one can determine the mass function for virialised dark matter haloes
well without including the baryons at least for the high mass end.

   At the same time as progress has been made studying the mass
function using n-body simulations, analytic formulae have been
developed to describe the mass functions.  The most influential work
was pioneered by \cite{PS74} which utilises the spherical top-hat
model of collapse.  Comparisons of the Press-Schechter mass function
to n-body simulations by a number of authors have noted that the
Press-Schechter mass function tends to overestimate the mass function
at low masses and underestimate the high mass end.  This discrepancy
has led to the development of improved analytic mass function
formulae, such as the work of \cite{ST99,SMT01,ST02} (S-T hereafter),
based on a model which accounts for ellipsoidal rather than spherical
collapse.  However in this paper we will concentrate on the
`universal' mass function formula presented in \cite{J01} (J01
hereafter), which is an empirical fit to a wide range of simulation
data (of which those used by S-T form a subset).  As we will show in
the next section this formula does actually predict the mass functions
in a SUGRA quintessence model without needing modification.

Both simulations and the analytic formulae show for CDM models with
Gaussian power spectra that the high mass end of the halo mass
function is very sensitive to the precise normalisation of the power
spectrum.  If one can by some means determine the halo mass function
over a range of redshifts then in principle one can find how the
linear growth factor evolves as a function of redshift for the
universe.

 The dark matter halo mass function is not directly observable, though
future observations exploiting weak gravitational lensing will make
progress in this direction.  However it is believed that the largest
dark matter haloes are recognisable because they host galaxy clusters.
Galaxy clusters contain, as well as galaxies, copious amounts of x-ray
emitting gas.  The presence of x-ray emitting gas not only signals the
presence of the cluster halo which helps to find and count them, but
can also be exploited to estimate the actual mass of the surrounding
dark matter halo in a variety of ways.  Estimates of the halo mass
function using observational data of galaxy clusters have been made and
utilised by many authors to estimate the value of the matter power 
quantity $\sigma_8$ locally -
e.g. \cite{WEF93},\cite{VL96}, \cite{Eke96}, \cite{seljak02}.
Observations of clusters over a range of redshifts potentially enables
the evolution of the growth factor to be determined.  
Two difficulties to surmount in using clusters as cosmological probes 
are the relation of the observables to theoretical characteristics, 
e.g.\ mass, and then the extraction of cosmological parameters free 
from astrophysical factors and degeneracies.  The typical
uncertainty in the theoretically determined mass functions are around
10\%. However making accurate mass estimates of galaxy clusters is far
from easy and comparing an observationally determined mass function
with the relatively clean theoretical mass function requires extensive
and detailed modelling - it is not our aim in this paper to discuss
these important aspects.

 As mentioned above the mass function formula given in J01 is a fit to
the mass functions from a range of CDM simulations.  These simulations
include models with a cosmological constant ($w=-1$) and Open models
which have an effective value of $w=-1/3$.  The values of interest for
$w$ for the dark energy are typically within the range that has
already been studied so it would not seem too surprising if the mass
function for models with constant values of $w$ intermediate between
$-1/3$ and $-1$ are accurately predicted by existing mass function
formulae.  However it is interesting to ask how the mass function of a
model where the value of $w$ is rapidly changing compares to the
predictions.  The SUGRA model of \cite{braxm} is an attractive model to
choose for this reason because the value of $w$ evolves reasonably rapidly.

  The part of the mass function which is most sensitive to the linear
growth factor is the high mass end where the mass function falls off
steeply.  One would similarly expect that it would be this region
of the mass function that would show the greatest sensitivity to a
time variation in $w$. We therefore have designed our simulation to
model a large volume of space so that we can measure the abundance of
very massive, but rare, clusters.

  As far as we are aware the only simulations with a quintessence
component with a changing EOS are those of \cite{Klyp03} who also
model the SUGRA quintessence together with a model by \cite{RP88}, which
has a much more gently changing value of $w$, and some models with
constant $w$.  Their simulations have very good spatial and mass 
resolution, allowing them
to study the properties of individual dark matter haloes in some
detail.  However their simulation cubes are relatively small compared to
ours: even their largest simulation cube is some 66 times smaller
in volume than our cube.  The results the authors report for the mass
functions determined from their simulations appear to be fully
consistent with our own results.  Our simulations are designed to give
better statistics at the high mass end of the mass function and are
in this sense complementary.

Below we describe the n-body code, the parameter choices for the
simulation and the initial conditions for the simulation. The results
themselves are reported in the section \ref{sec.res}.

\subsection{Code details.}

  We have used the publicly available parallel code GADGET (v1.1)
described in \cite{Spr01} to perform the n-body simulation.  We
modified the code to allow the inclusion of a quintessence component
with an EOS of the form $w(a) = w_0 + (1-a)w_a$, where $a$ is
the expansion factor defined so that $a=1$ at the present epoch.  The
only modifications required to the code were 
to update expressions for the Hubble parameter to include the quintessence 
component. We verified these modifications
by checking that the modified code can reproduce the correct
linear growth rates on a series of test simulations.
We will only report the results of this test for the simulation
presented here.

 We determined the linear growth rates from the n-body simulation by
measuring the power associated with the longest wavelength modes of
the simulation box for the initial conditions and for each output.
The longest wavelength modes are least affected by non-linear effects.
For comparison we also computed the growth rates by numerical
integration of equation (\ref{eq.ga}). Between z=38.9 and z=2 the 
agreement in 
the growth factors measured from the simulation and computed by
numerical quadrature is better than 0.2\%. This is a non-trivial test
that the code can handle a quintessence component because the SUGRA 
model has an EOS $w=-0.25$ at the upper redshift and by
redshift 2 the quintessence component already has a significant 
dynamical influence: for
example the value of $G(a)$ is 0.84 rather than unity as it would be in
a matter dominated universe. The long wavelength modes grow slightly
more slowly than the linear theory prediction, due we believe to
non-linear effects \citep[see][]{BE94}.
By z=0, when the long wavelength density fluctuations themselves have
an amplitude of a few percent, the agreement is better than 0.8\%.  By
comparison this difference is 25 times smaller than the difference of the
growth rates of the quintessence model and \lcdm\ over the same
interval.

\subsection{Simulation details.}
 
  We have designed the simulation with two aims in
mind:$\phantom{xxx}$ (i) to determine the high end of the halo mass
function, and (ii) to match as closely as possible, at redshift zero, the
parameters of an existing \lcdm\ simulation so that a direct
comparison can be made.  The most suitable reference \lcdm\ simulation
for determining the high mass end of the halo mass function is the
\lcdm\ Hubble volume run described in \cite{Evrard02} which models a
cube, $3000\hmpc$ on a side.  We have therefore selected the same
values for the matter density, $\Omega_m$, the particle mass and the
gravitational softening length as the \lcdm\ Hubble simulation. We
also match very closely the power spectrum shape and amplitude of the
linearly extrapolated power spectrum at redshift zero to those of the
\lcdm\ Hubble simulation.  Our simulation volume is significantly
smaller than the \lcdm\ Hubble simulation at $648\hmpc$ on a
side.

 We have chosen a quintessence component with a variable value
of $w$.  As discussed earlier, the choice $w_0=-0.82$, $w_a=0.58$,
gives a good fit to the SUGRA quintessence model. We
will refer to our simulation as the SUGRA-QCDM model. The parameters of the
simulation are listed in table \ref{params}.

 The initial conditions were generated by the serial version of the
code that was used to generate the initial conditions for the Hubble
volume simulations \citep{Evrard02}. The initial conditions are
created from an initially uniform particle distribution 
\citep[in this case
a glass distribution generated in the way described by][]{White96} by
perturbing the particles to give the desired power spectrum and using
the Zel'dovich approximation \citep{Zeld} to assign velocities to
each particle which are proportional to the displacements.  The code
needed to be altered to give the correct constant of
proportionality, which depends on the logarithmic growth 
${\rm d}\ln D/{\rm d}\ln a$, when
assigning the velocities.  The difference in this constant between
an Einstein-deSitter ($\Omega_m=1$) and quintessence model at the start 
redshift is small - 
only $\sim1.5\%$ - though this is much larger than the difference 
between a matter dominated and cosmological constant model.

\begin{table}
\caption{Parameters of the SUGRA-QCDM n-body simulation - the mass resolution,
softening and power spectrum normalisation are chosen to match
the \lcdm\ Hubble volume simulation described in Evrard et al 2002.
$\sigma_8$ is the RMS of the linear density field smoothed with a 
top-hat filter of $8\hmpc$ in radius.   }
\label{params}
\begin{tabular}{@{}lcr}
\hline
Parameter & & Value  \\
\hline
$\Omega_m$          &$\phantom{xxxx}$ &  0.30 \\
$w_0$               &                           & -0.82 \\
$w_a$               &                           &  0.58 \\
$\sigma_8(z=0)$          &                           &  0.90 \\
Box size/$\hmpc$    &                           &   648 \\
Number of particles &                        & 10077696\\
Particle mass/$10^{12}\hmsun$ &       &            2.25\\
Comoving softening length/$\hmpc$     &           & 0.1 \\
\hline
\end{tabular}
\end{table}

\section{Simulation Results}\label{sec.res}
  The simulation was started at z=38.9 and outputs were made at redshifts
z = 3, 2, 1.5, 1, 0.5, 0.25 and 0.   

\subsection{The mass function at z=0}

 To begin with we will compare the mass functions at redshift zero.
The z=0 linear power spectrum of the SUGRA-QCDM simulation has
the same shape and amplitude as the \lcdm\ Hubble
simulation by design.  Analytical fitting formulae for mass functions 
such as \cite{PS74}, S-T and \cite{J01},
all predict that the CDM mass function depends
primarily on the linear power spectrum.  We would therefore expect
if one extrapolates these results to quintessence models that our
two models should have very similar mass functions at 
redshift zero.

  One potential complication in comparing two different cosmological
models lies in the problem of how to define the haloes in a consistent
way so that they can be compared.  A common approach to this problem is to use
the spherical top-hat collapse model to give guidance as to the
expected collapse overdensity of virialised objects - e.g. \cite{LC93,Eke96}.
Following these arguments would lead to different values for 
the overdensity for haloes selected in \lcdm\ and SUGRA-QCDM.

 However we can circumvent these issues by adopting the halo definition
used in J01.  The haloes are defined using the friends-of-friends algorithm 
\citep{DEFW} with a linking length of b=0.2. J01 showed that using 
this way of defining haloes it is possible to fit the mass functions for CDM
models with a wide range of cosmological parameters and 
redshifts with a single `universal' fitting formula which is 
accurate to better than 20\%.  It is this simplicity which
motivates our choice of halo definition. In this subsection we will 
compare the SUGRA-QCDM and \lcdm\ simulations at z=0 directly.

   Our ability to determine the mass function is limited by a number
of considerations.  At the high mass end, where the objects themselves
are resolved with the largest numbers of particles, the main limitation in
determining the mass function is the volume of the box.  We plot the
mass functions only up to the point where the Poisson error reaches
10\%.  At the low mass end the situation is less simple.  One would
expect determination of the mass function to be less and less reliable
as one resolves haloes with fewer and fewer particles.  We take a
limit of 20 particles as the minimum number of particles.  This limit
is supported by tests presented in appendix A of J01 where they
compare the mass functions from simulations with the same mass
resolution as our SUGRA-QCDM simulation to those from simulations
with superior mass resolution.

\begin{figure}
\includegraphics[width=8cm]{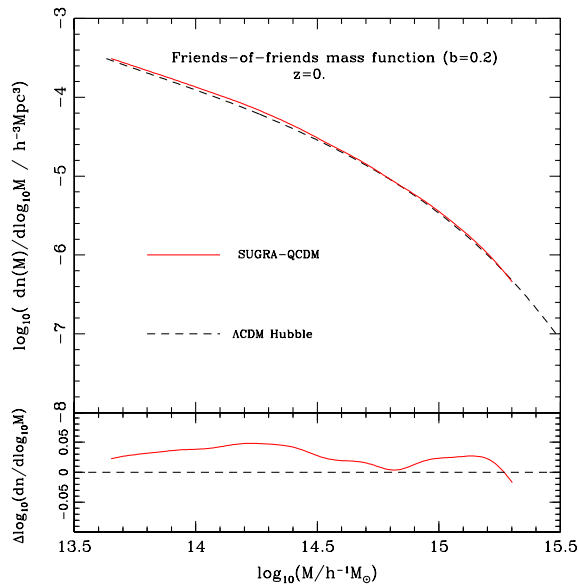}
\caption{Top panel: the dark matter halo mass function defined using
  the friends-of-friends algorithm with a linking length of 0.2.  The
  mass functions of \lcdm\ Hubble simulations and SUGRA-QCDM that  
  at z=0 have nearly identical power spectra are very
  similar.  The mass functions are plotted for haloes of 20 particles
  and more and up to an upper mass cut-off where the
   Poisson uncertainty in the mass function first exceeds 10\%.  
  Bottom panel:  the solid curve shows the difference between
  the SUGRA-QCDM mass function and the \lcdm\ Hubble volume mass function.
  The differences are below 10\%.}
\label{z0_massfn}
\end{figure}

In figure \ref{z0_massfn} we show the SUGRA-QCDM and \lcdm\ mass
function evaluated with the FOF(b=0.2) group finder.  The mass
functions from both simulations have been smoothed using an identical
Gaussian with an RMS width of 0.08 dex. The mass functions agree very
well - the differences are below 10\%.  This is an encouraging result
\citep[ a similar result is also seen in ][]{Klyp03} and suggests that
it should be possible to predict the mass functions for this and other
quintessence models with a smoothly varying value of $w$ with a
precision of $\sim10$\% from a knowledge of just the linearly evolved
power spectrum.  In the next section we will compare the SUGRA-QCDM
mass function with the universal mass function formula of J01 directly
for several redshifts.

\subsection{The mass function at all redshifts}

  Here we will plot the combined mass functions at
redshifts 3, 2, 1.5, 1, 0.5, 0.25 and 0 for the SUGRA-QCDM run and
compare them with the J01 mass function
formula.  But we first repeat a few definitions, adapted from J01,
which are required further on.

  It is convenient to use as an effective mass variable,
$\ln\sigma^{-1}$, where $\sigma(M)$ is the RMS of the linear density
field smoothed with a top-hat filter containing mass $M$ at the mean
density. This is defined as:

 \begin{equation}\label{vardeff}
\sigma^2(M,z)  =  {D^2(z)\over2\pi^2}\int_0^\infty k^2P(k)W^2(k;M){\rm d}k,
\end{equation} 
where $D(z)$ is the growth factor of linear perturbations normalised
so that $D(0)=1$, $P(k)$ is the linear power spectrum at
redshift zero and $W(k;M)$ is the Fourier-space representation of a
real-space top-hat filter enclosing mass $M$ at the mean density of
the universe.

We define the mass function $f(\ln\sigma^{-1}, z)$ through:
\begin{equation}\label{deff}
   f(\ln\sigma^{-1}, z) \equiv \frac{M}{\rho_0}{{\rm d}n(M, z)\over{\rm d}\ln\sigma^{-1}},
\end{equation}
where $n(M, z)$ is the abundance of halos with mass less than $M$ 
at redshift $z$, and $\rho_0(z)$ is the mean density of the universe at 
that time.  With these definitions we can plot the mass functions for
any CDM model and at any redshift conveniently onto the 
$\ln f - \ln\sigma^{-1}$ plane.  

\begin{figure}
\includegraphics[width=8cm]{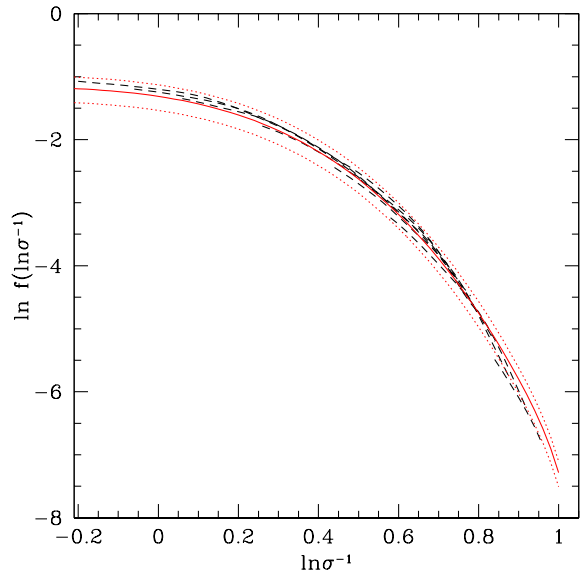}
\caption{ The halo mass functions for the SUGRA-QCDM simulation plotted 
in the $f-\ln\sigma^{-1}$ plane for redshifts 0,0.25,0.5,1,2,3.  
The smooth solid curve flanked by dotted
curves marks the J01 universal mass function formula. The
dotted lines mark a  $\pm20$\% uncertainty in the mass predicted mass function.
The SUGRA-QCDM mass functions are plotted for groups of 20 or more particles
and up to the point where the Poisson error becomes 10\%.
The SUGRA-QCDM mass function is well fit by the Jenkins et al formula.
Some departure is apparent at high values of $\ln\sigma^{-1}> 0.85$, but
as discussed in the text the origin of this departure is likely due to
numerical limitations rather than being real. }
\label{all_massfn}
\end{figure}

  Figure \ref{all_massfn} shows the mass functions for the 
SUGRA-QCDM simulation, plotted in the $\ln f - \ln\sigma^{-1}$ plane, 
for redshifts $z$= 0, 0.25, 0.5, 1.0,
1.5, 2 and 3.   The mass functions are plotted as dashed lines 
for haloes of 20 particles and more (as in J01) and cut-off at 
the high mass end when the Poisson errors reach 10\%.   The simulation curves 
have been smoothed using a Gaussian with an RMS of 0.05 dex.  The solid line
shows the J01 universal mass function fit given by:
\begin{equation}
\label{tcdm_fit2}
 \phantom{xxxxxxx}f(\ln\sigma^{-1}) =
 0.315\;\exp\big[-|\ln\sigma^{-1}+0.61|^{3.8}\big],
\end{equation}
valid over the range $-1.2\le\ln\sigma^{-1}\le1.05$,
and the flanking dotted lines denote a 20\% uncertainty about this fit. 
These curves are similarly convolved with a 0.05 dex Gaussian.  

Note that the universal nature does not imply that different models predict 
the same observable consequences, only that they can be treated 
by the same parametrization.  Each model is distinguished by its 
particular $\sigma^2(M,z)$ relation. 

  The main result of this section is that the SUGRA-QCDM mass function
is well fit by the J01 formula - at least to the same degree of
accuracy as was found for a host of other CDM cosmological models.
The fit is good despite the fact that this model has a changing value
of $w$.  It therefore seems likely that the J01 should work well for a
wide range of dark energy models.  Although we have used the J01
formula in figure \ref{all_massfn} we could instead have used the formula of
S-T.  \cite{Klyp03} find that the
S-T formula provides an excellent fit to their mass functions.
We do find however that for $\ln\sigma^{-1}$ above about 0.75
that J01 formula is definitely a better fit, with the S-T formula tending
to overpredict the number of haloes. 

  It is apparent from figure \ref{all_massfn} that the most
significant deviation away from J01 formula occurs for
$\ln\sigma^{-1}>0.85$, where only the $z=3$ output contributes.  A
closer analysis of this curve and other curves shows that the plotted
mass functions tend to drop away relative to the J01 curve for small
numbers of particles per group. This drop off is particularly apparent
for the $z=$ 3 and 2 outputs where the mass function is steep.  The
cause of this feature is unclear but is likely numerical in origin.
The influence of the dark energy component diminishes as one goes to
higher redshift and the J01 formula works well for matter dominated
models - so if anything one would expect the degree of fit to improve
with increasing redshift contrary to what is seen.  This divergence is
much less pronounced at lower redshift and is not apparent in the
resolution tests presented in appendix A of J01 for the Hubble volume
simulations which are for redshift zero.

  To conclude, firstly we have found in this section that the $z=0$
mass functions of the Hubble \lcdm\ simulation and our own SUGRA-QCDM
simulation match to a precision of better than 10\% in abundance for
haloes in the mass range of $4.5\times10^{13}-2\times10^{15}\hmsun$.
By design, the $z=0$ linear power spectrum of the two models was
closely matched. The results indicate that the mass function depends
primarily on the linear power spectrum and is only very weakly if at
all dependent on the details of the expansion history.  Secondly, from
an analysis of the SUGRA-QCDM mass function at a range of redshifts we
find that mass function formulae such as J01 and S-T provide good
estimates of the halo mass function for flat quintessence models. The
J01 formula gives a better fit at the high mass end than S-T for
$\ln\sigma^{-1}>0.75$.  The goodness of fit is comparable with that
found when comparing these formulae to the mass functions determined
from the following cosmological models: matter dominated, open, and
flat models with a cosmological constant.  We are able to verify
the accuracy of the fit over range $-0.2<\ln\sigma^{-1}<0.85$ from our
simulation. 

 These results imply that the main observational discriminators
for large-scale structure between cosmological models are: the present
linear power spectrum, to be fixed by wide field surveys; the linear
growth factor, discussed in the first part of this paper and probed by
future deep surveys; and distances and volumes measured by expansion
history mappers such as SNAP.

\section{Conclusions}\label{sec.concl}

Cosmic structure formation and evolution provides an 
additional path to exploring the cosmological model besides 
mapping the expansion history.  Many of the observational probes 
employing structure rely on the growth dynamics of density 
fluctuations, in either the linear or nonlinear regime. 

Linear perturbations offer the best hope of structure measurements 
with relatively uncomplicated physics and clean observations free of 
many astrophysical entanglements.  Analyzing the linear growth 
behavior, we find little leverage on the nature of the dark 
energy -- unless it possesses a time varying equation of state. 
In this latter case the growth shows good complementarity with 
supernova and CMB measurements.  

We have emphasized that such time variation appears generically 
in proposed dark energy models other than the cosmological constant 
and can also be used to treat modifications of the cosmological 
framework. 

The growth also enters as a 
contribution to the gravitational lensing power spectrum, and 
weak lensing as a cosmological probe promises interesting 
complementarity and crosschecks.  Such a method can take 
advantage of both the linear and nonlinear density scales to 
attempt to balance sensitivity with systematic uncertainties. 

For the nonlinear realm of structure, investigations at the level 
of precision necessary require numerical simulations.  We present 
one of the first, and currently the largest volume, N-body simulation 
with a time varying equation of state.  As a first application of 
the data, we calculate the halo counts as a function of mass, a 
quantity relevant to forthcoming structure surveys.  We find that 
this is well fit by the previous Jenkins mass formula, extending 
its universality to time varying equations of state, at least at 
the 20\% level.  
At redshift zero the dynamical dark energy model 
shows agreement to better than 10\% with the cosmological constant 
simulation that matches the linear power spectrum today, suggesting 
this serves generally as a central 
quantity in describing structure formation. 

Much work remains for the future to bring observations of structure 
to the level of theoretical and systematic uncertainty necessary 
for precision probing of dark energy.  But the results here lay 
a foundation to build upon for using not only the expansion 
history of the universe but the growth history of structure in a 
rigorous quest to understand the physics behind the accelerating 
universe.

\section*{Acknowledgments}
We wish to thank Carlos Frenk for encouragement and Dragan Huterer 
and Bhuvnesh Jain for useful discussions.  EL acknowledges 
support for this work from the Director, Office of Science, US DOE 
under DE-AC03-76SF00098 at LBL.  EL would also like to thank the 
University of Massachusetts and University of Pennsylvania for 
hospitality during part of the paper preparation. 


\section*{Appendix. Approximation of the Linear Growth Factor}
\label{sec.appx} 

Within the formalism of the Birkhoff's theorem argument 
presented by \cite{pje80,pje93} for the evolution of 
linear density perturbations, one can write a closed form 
expression for the growth.  However this does not hold 
for general cosmological models.  While it was sufficient 
and indeed prescient for the main models considered in 1980, 
it neglects a term arising for a general pressure component 
(as Peebles alludes to); 
this term is proportional to $(1+3w)(1+w)\Omega_w$ and so we 
see that serendipitously the closed form is exact for a pure 
matter model (SCDM, $\Omega_w=0$), an open model (OCDM, 
$w=-1/3$), and a cosmological constant model ($\Lambda$CDM, 
$w=-1$).  But a second order differential equation does not 
generally possess such a quadrature and instead an equation 
like Eq.\ (\ref{da}) must be solved. 

\begin{figure}
\includegraphics[width=8cm]{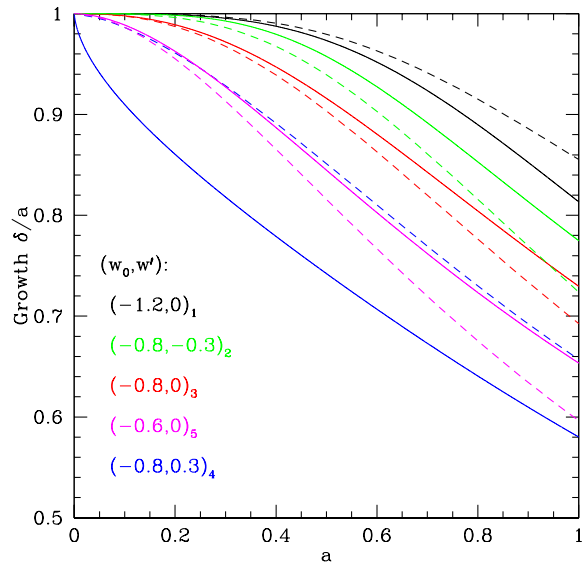}
\caption{Comparison of the exact growth factor (solid curves) 
with the closed form approximation, Eq.\ \ref{closed} (dashed 
curves).  The solid curves are labeled according to the 
vertical ordering, and the subscripts indicate the ordering of 
the dashed curves for those models.  The approximation is 
insufficient for precision cosmology. 
} 
\label{fig.lamgrow}
\end{figure}

Fig.\ \ref{fig.lamgrow} shows the difference between the 
exact solution for the growth factor and the approximation 
given by the closed form 
\beq 
G_{approx}={5\over2}\om{H\over a}\int_0^a da\,(aH)^{-3}\,.
\label{closed}\eeq 
The differences for the time varying SUGRA model can be 
$\sim15\%$; this propagates into the power spectrum as  
the square, and then further as a possibly substantial bias 
on the cosmological parameters.  The situation is 
better if future detailed observations allow use of the 
growth factor normalized to the present ($\sim1\%$ difference 
for SUGRA, though $\sim5\%$ for $w=-1.2$).  See \cite{wangste} 
for a fitting function involving an integral over the equation 
of state, valid at high redshifts, for slow variations. 

\label{lastpage}

\end{document}